\documentclass[10pt,aps,prl,twocolumn,eqsecnum,amssymb,amsmath,showpacs,a4paper,superscriptaddress,nofootinbib,floatfix]{revtex4-1}
\usepackage{graphicx}
\usepackage{amsfonts}
\usepackage{amsmath}
\usepackage{hyperref}
\usepackage{units}
\usepackage[latin1]{inputenc}      		
\usepackage{color}
\usepackage{dcolumn}
\usepackage{bm}
\usepackage{cleveref}
\usepackage{braket}
\crefname{section}{§}{§§}
\Crefname{section}{§}{§§}
\usepackage{mathtools}
\usepackage{chngcntr}
\counterwithout{equation}{section}
\usepackage[normalem]{ulem}
\usepackage{url}

\newcommand{\be}{\begin{equation}}
\newcommand{\ee}{\end{equation}}
\newcommand{\bsub}{\begin{subequations}}
\newcommand{\esub}{\end{subequations}}
\newcommand{\bea}{\begin{eqnarray}}
\newcommand{\eea}{\end{eqnarray}}
\newcommand{\bi} {\begin{itemize}}
\newcommand{\ei} {\end{itemize}}
\newcommand{\bmat} {\begin{pmatrix}}
\newcommand{\emat} {\end{pmatrix}} 

\newcommand{\D}{\mathrm{d}}
\newcommand{\I}{\mathrm{i}}
\newcommand{\E}{\mathrm{e}}
\newcommand{\op}[1]{\ensuremath{\boldsymbol{#1}}}
\newcommand{\lagrange}[1]{\mathcal{L}_{\text{#1}}}

\makeatletter
\newcommand*{\balancecolsandclearpage}{%
  \close@column@grid
  \clearpage
  \twocolumngrid
}
\makeatother

\makeatletter
\let\cat@comma@active\@empty
\makeatother

\begin{document}

\title{Interferometric Unruh Detectors for Bose-Einstein Condensates}

\author{Cisco Gooding}
\affiliation{School of Mathematical Sciences, University of Nottingham, University Park, Nottingham, NG7
2RD, UK}

\author{Steffen Biermann}
\affiliation{School of Mathematical Sciences, University of Nottingham, University Park, Nottingham, NG7
2RD, UK}

\author{Sebastian Erne}
\affiliation{School of Mathematical Sciences, University of Nottingham, University Park, Nottingham, NG7
2RD, UK}
\affiliation{Vienna Center for Quantum Science and Technology, Atominstitut, TU Wien, Stadionallee 2, 1020 Vienna, Austria}
\affiliation{Wolfgang Pauli Institut, c/o Fak. Mathematik, Universit{\"a}t Wien,
Nordbergstrasse 15, 1090 Vienna, Austria}

\author{Jorma Louko}
\affiliation{School of Mathematical Sciences, University of Nottingham, University Park, Nottingham, NG7
2RD, UK}

\author{William G. Unruh}
\affiliation{Department of Physics \& Astronomy, University of British Columbia, Vancouver, BC V6T 1Z1, Canada}
\affiliation{Hagler IAS, IQSE, Texas A\&M, College Station, TX, 77843-4242, USA}

\author{J\"org Schmiedmayer}
\affiliation{Vienna Center for Quantum Science and Technology, Atominstitut, TU Wien, Stadionallee 2, 1020 Vienna, Austria}

\author{Silke Weinfurtner}
\affiliation{School of Mathematical Sciences, University of Nottingham, University Park, Nottingham, NG7
2RD, UK}
\affiliation{Centre for the Mathematics and Theoretical Physics of Quantum Non-Equilibrium Systems, University of Nottingham, Nottingham, NG7 2RD, UK}

\date{\today}

\begin{abstract}
The Unruh effect predicts a thermal response for an accelerated detector moving through the vacuum. Here we propose an interferometric scheme to observe an analogue of the circular Unruh effect using a localized laser coupled to a Bose-Einstein condensate (BEC). Quantum fluctuations in the condensate are governed by an effective relativistic field theory, and as demonstrated, the coupled laser field acts as an effective Unruh-DeWitt detector thereof. The effective speed of light is lowered by $12$ orders of magnitude to the sound velocity in the BEC. For detectors traveling close to the sound speed, observation of the Unruh effect in the analogue system becomes experimentally feasible.
\end{abstract} 

\maketitle


\textbf{\textit{Introduction.}---} The formulation of quantum field theory (QFT) in curved spacetime highlights the ambiguity of the definition of particle and ``vacuum'' state which persists even in flat spacetime. A prominent example thereof is the Unruh effect \cite{unruh76} (for a pedagogical introduction see e.g.~\cite{ben2019unruh}). Originally stated it predicts that a uniformly linearly accelerated observer sees fluctuations of the Minkowski vacuum as a thermal bath with a characteristic temperature 
\begin{equation}\label{eq:T_unruh}
k_\mathrm{B} T_\mathrm{U} = \frac{\hbar\,a}{2\pi\,c} ~,
\end{equation}
proportional to its acceleration $a$. Direct experimental verification is however to date still missing. The main problem is that the Unruh temperature \eqref{eq:T_unruh} is inversely related to the propagation speed $c$ of the field. Hence very large accelerations are required to produce a measurable temperature for fundamental quantum fields. 

Analogue (quantum) simulators enable the study of relativistic QFT effects in well controlled laboratory setups \cite{LivingReview,volovik2003universe,PhysRevLett.91.240407}. In these analogue systems, the speed of sound replaces the speed of light for the propagation speed of the effective massless field describing the evolution of perturbations in the system. For an experimental observation see e.g.~\cite{langen2013local}. This enabled a number of first experimental observations of fundamental effects of QFT in curved spacetime, like e.g.~superradiant scattering/amplification from a rotating analogue black hole (dumb hole) \cite{torres2017rotational}, the analogue of the Hawking effect \cite{Hawking} in classical \cite{weinfurtner11,classAn,Philbin1367} and quantum systems \cite{PhysRevLett.105.203901,Steinhauer,quantAn}, cosmological particle production \cite{PhysRevLett.99.201301}, inflationary scenarios \cite{PhysRevX.8.021021}, and the dynamical Casimir effect \cite{PhysRevLett.109.220401}.

In this letter we address the question if the same can be done for the Unruh effect. Can one ``accelerate'' a detector in the vacuum state of some field and see a thermal response according to Eq.~\eqref{eq:T_unruh}? Several theoretical proposals can be found in the literature \cite{PhysRevA.95.013627,PhysRevLett.101.110402,PhysRevLett.97.190405,PhysRevLett.94.220401,adjei2020quantum}, though experimental efforts to observe the analogue Unruh effect have relied on either functional equivalence \cite{Hu2019} or virtual observers \cite{PhysRevA.98.022118}. In contrast, we propose a physically accelerated particle detector, constructed using a localized laser beam interacting with an oblate quasi-2d Bose-Einstein condensate (BEC). Long-wavelength density perturbations in a homogeneous BEC are described by an effective relativistic field theory which transduces fluctuations in the refractive index of the medium into phase fluctuations in the laser. Using an interferometric setup, we show that the laser, or any continuous probing field, realizes a suitable particle detector and demonstrate that for an accelerated circular path of the laser-BEC interaction point, in the (Minkowski) vacuum of the density perturbation field, the effect of the Unruh temperature can indeed be measured.

\textbf{\textit{Circular Unruh effect with a transverse detector field.}---}Bell and Leinaas \cite{bell83,bell87} (see also Unruh \cite{unruh98}) showed that circular motion with its constant radial acceleration could also produce a spectrum which was approximately thermal. The advantages of uniform circular motion for analogue relativistic field theories have been previously acknowledged \cite{PhysRevLett.101.110402}. A notable simplification that occurs for uniform circular motion is that the proper and coordinate time are related by a time-independent gamma factor. We first consider an idealized field theory, and present a new demonstration that a relativistic field can serve as a detector for the circular Unruh effect.

The field theory involves a two-dimensional scalar field $\phi(t, \op{x})$, with $\op{x}=(x,y)$, and a one-dimensional scalar probing field $\psi(t,z)$. The Lagrangian is written as
\begin{align}
L ~=~ & \frac{1}{2}\int\!\D z \left(\dot{\psi}^2(t,z)-\left(\partial_z\psi(t,z)\right)^2\right) \nonumber\\
& +\frac{1}{2}\int\!\D \op{x}\left(\frac{1}{c_s^2}\dot{\phi}^2(t,\op{x})-\left(\nabla\phi(t,\op{x})\right)^2\right) \nonumber\\
& -\varepsilon \int\!\D \op{x}\D z\;\partial_t\psi(t,z)\phi(t,\op{x})\delta(\op{x}-\op{X}(t))\delta(z),
\label{eq:Eff}
\end{align}
where $c_s$ is the propagation speed of the $\phi$ field, $\bm{X}(t)$ parametrizes the path of interaction between the two fields, and $\varepsilon$ is a small coupling constant. The propagation speed of the $\psi$ field has been set to unity. The delta functions $\delta(\bm{x}-\bm{X}(t))\delta(z)$ restrict the interaction to a trajectory of the $\psi$ beam in the $z=0$ plane described by $\bm{X}(t)$, such that the effective interaction Lagrangian is
\begin{equation}
L_\text{int}=-\varepsilon \phi(t,\bm{X}(t))\partial_t\psi(t,0).
\label{eq:Lint}
\end{equation}

The equation of motion for $\psi$ reads
\begin{equation}
\partial_t^2\psi(t,z) - \partial_z^2\psi(t,z)~=~ \varepsilon \delta(z) \partial_t\phi(t, \op{X}(t)).
\end{equation}
This equation has the approximate solution
\begin{equation}\label{psisol}
\psi(t,z) ~=~ \psi_0(t,z) + \frac{\varepsilon}{2} \phi\left(t-\left|z\right|, \op{X}\left(t-\left|z\right|\right)\right),
\end{equation}
which shows that the field $\phi$ leaks into the probing field $\psi$. In other words, the interaction acts as a transducer from the $\phi$ field to the $\psi$ field. At the same time, fluctuations of $\psi_0$ will leak into the $\phi$ field as a backaction.

We take the interaction trajectory to be circular with radius $R\ge0$ and angular frequency $\Omega\ge0$, such that $\op{X}(t)=(R\cos(\Omega t),R\sin(\Omega t),0)$. In the rest of this section we consider a toy scenario in which $\psi$ is initially prepared in its vacuum state.

Suppose hence that prior to the interaction $\phi$ is in the vacuum state $\ket{0}$ and the detector field $\psi$ is in its ground state $\|0\rangle\!\rangle$. First-order transitions connect the initial state $\ket{0}\|0\rangle\!\rangle$ to states of the form $\ket{1_{\op{k}}}\|1_{K}\rangle\!\rangle$ where $\ket{1_{\op{k}}}=a_{\op{k}}^\dagger\ket{0}$ and $\|1_{K}\rangle\!\rangle=\bar{a}_{K}^\dagger\|0\rangle\!\rangle$, with $a_{\op{k}}^\dagger$ and $\bar{a}_K^\dagger$ being the creation operators in the mode expansions of $\phi$ and $\psi$, respectively. The transition amplitude is 
\begin{align}
&\I\varepsilon \int\!\D t\,\langle\!\langle
1_K\|\langle 1_{\bm{k}}|\partial_t\psi(t,0)\phi(t,\bm{X}(t))\ket{0}\|0\rangle\!\rangle
\notag
\\
&= 
\frac{\varepsilon \sqrt{\tilde\omega/\omega}}{2 {(2\pi)}^{3/2}}
\int\! \D t \,\E^{\I\tilde{\omega} t}\E^{-\I\bm{k}\cdot\bm{X}(t)+\I\omega t},
\label{TA2}
\end{align}
where $\tilde{\omega}=K>0$ is the detector mode frequency and $\omega=c_s |\bm{k}|$ is the $\phi$ mode frequency. Note from (\ref{TA2}) that the transition amplitude vanishes in the special case of a static trajectory, just as for a pointlike relativistic Unruh-DeWitt detector on an inertial trajectory \cite{BD}.

Taking the squared modulus of (\ref{TA2}), summing over all possible final $\phi$ states, and using the stationarity of the circular trajectory to factor out the (formally infinite) total observation time, we find that the transition probability per unit time is 
\begin{equation}\label{TP}
\frac{\varepsilon^2 \tilde\omega}{2\pi} 
\int\!\D s \, \E^{-\I\tilde{\omega}s}W(s),
\end{equation}
where $W(s)$ is the $\phi$ field Wightman function evaluated on the interaction trajectory, 
\begin{equation}\label{Wightman}
W(s)=\langle 0 |\phi(s,\bm{X}(s))\phi(0,\bm{X}(0))|0\rangle . 
\end{equation}
The transition probability per unit time has thus the same dependence on the interaction trajectory as that of a pointlike two-state system coupled to $\phi$ along the trajectory \cite{BD}: \emph{the $\psi$ field acts as a detector for fluctuations of the $\phi$ field along the interaction trajectory\/}. Note, however, that $\tilde\omega$ here is the energy with respect to the laboratory time~$t$, while the normal Unruh effect context is for trajectories satisfying the timelike condition $\Omega R < c_s$, and for energies defined with respect to the proper time $\tau = t/\gamma_s$, where $\gamma_s = \left(1-(\Omega R)^2/c_s^2\right)^{-\nicefrac{1}{2}}$. This gamma-factor will be crucial when estimating the experimental feasibility for detecting the analogue circular Unruh effect.

\textbf{\textit{Lasers as local detectors of a BEC field.}---}In the following, we establish the connection between the idealized field theory model, Eq.~(\ref{eq:Eff}), and a localized laser beam propagating in the $z$-direction interacting with an effectively two-dimensional BEC in the $(x,y)$-plane.

The free electromagnetic field Lagrangian in $(3+1)$ dimensions is $\lagrange{em}=-\frac{1}{4}F_{\mu\nu}F^{\mu\nu}$, with field tensor $F_{\mu\nu}=\partial_\mu A_\nu-\partial_\nu A_\mu$ and vector potential $A_\mu$. In the Coulomb gauge ($A_t=0$ and $\nabla\cdot \bm{A}=0$), for a linearly polarized laser propagating in the $z$-direction, perpendicular to the BEC plane, the Lagrangian reduces to 
\begin{equation}
\lagrange{em}=\frac{1}{2}\left(\left(\partial_t A(t,z)\right)^2-\left(\partial_z A(t,z)\right)^2\right),
\label{eqn::Lem}
\end{equation}
with the speed of light set to unity. As the laser is moved, the interaction point traces out a path $\bm{X}(t)$ in the $(x,y)$-plane. We express the evolution in terms of the laboratory time $t$. As before, we specialize to uniform circular trajectories, in which case the gamma factor $\gamma_s=t/\tau$ is constant.

The homogeneous quasi-2d BEC is described by the Lagrangian (see e.g.~\cite{PethickSmith})
\begin{equation}
\lagrange{BEC}=\I\hbar\Phi\partial_t\Phi^*+\frac{\hbar^2}{2m}\left|\nabla\Phi\right|^2+\frac{g_\text{2d}}{2}\left|\Phi\right|^4,
\label{eqn::Lbec}
\end{equation} 
where $\Phi=\Phi(t,\bm{x})$ is the complex-valued BEC field and $m$ is the boson atom mass. The two-dimensional interaction coupling constant is given by
\begin{align}
g_{2\mathrm{d}} = \sqrt{8\pi}\frac{\hbar^2 a_s}{m a_\perp} ~,
\end{align}
where $a_s$ is the s-wave scattering length, and $a_\perp=\sqrt{\hbar/m\omega_\perp}$ is the oscillator length of the transverse confinement $V(z) = m \omega_\perp^2 z^2/2$. The chemical potential $\mu=\rho_0g_{2\mathrm{d}}$ is given in terms of the coupling constant and the average density $\rho_0=\frac{N}{V}$ of $N$ bosons in a volume $V$ and defines the speed of sound in the BEC $c_s=\sqrt{\mu/m}$ and the healing length $\xi = \hbar/\sqrt{2\mu m}$. The effective two-dimensional BEC description applies to the regime $\mu, \, k_\mathrm{B}T \ll \hbar \omega_\perp$, for which dynamics along the transverse direction is frozen. Note that here we neglect swelling of the condensate in the $z$-direction, which, for the linearized equations of motion, only causes a slight shift in the speed of sound \cite{Salasnich}.

When the laser beam passes through the BEC, the atoms will react by forming dipoles according to their polarizabilities $\alpha$. Assuming the laser is sufficiently detuned from atomic resonance, $\alpha$ can be taken to be real and the BEC-light interaction can be calculated within a semiclassical model in the framework of macroscopic electrodynamics. In the dilute gas regime, $\alpha \rho_\mathrm{3d} \ll 1$, the Lagrangian describing the interaction is \cite{Cattani10,Pitaevskii61}
\begin{equation}
\mathcal{L}_{\text{int}} ~=~ \alpha\left(\partial_t A\right)^2\left|\Phi\right|^2 ~.
\label{eqn::Lint}
\end{equation} 
The BEC thickness $\Delta z$ is assumed to be small, as is the width of the laser beam, such that the interaction between the laser and the BEC is pointlike. Although not explicitly written everywhere, the fields $A$ and $\Phi$ in this interaction term are understood to be evaluated at $z=0$. The coupling Eq.~(\ref{eqn::Lint}) can be interpreted in terms of a fluctuating index of refraction $n_\mathrm{BEC} = \sqrt{1+\alpha \rho_\mathrm{3d}}$, which, within the dilute gas approximation, can be expanded to first order in $\alpha \rho_\mathrm{3d}$. The imaginary (absorptive) component of $\alpha$ follows from the Kramers-Kronig relation, and will be used in the following section to ensure the laser-BEC interaction is non-destructive.

In order to map the system to the idealized model Eq.~(\ref{eq:Eff}), we write the fields $A(t,z)$ and $\Phi(t,\op{x})$ in terms of perturbations about a classical background field ansatz, such that $A(t,z)=A_0\cos\left(\omega_L t-Kz+\psi(t,z)\right)$ and $\Phi(t,\op{x})=\sqrt{\rho_0+\delta \rho(t,\op{x})}\E^{\I \theta(t,\op{x})}$. We take the background laser field to be a plane wave, $A_0\cos\left(\omega_L t-Kz\right)$, and the perturbation $\psi(t,z)$ to be a real field that describes the phase fluctuations of the laser. Neglecting absorption, the amplitude of this plane wave can be taken to be constant. The BEC field $\Phi$ is expanded in terms of the (inhomogeneous) phase and density perturbations ($ \theta(t,\op{x})$ and $\delta \rho(t,\op{x})$, respectively) about the homogeneous mean field density $\rho_0 = \langle |\Phi|^2\rangle$. For the remainder of the paper, we consider the BEC perturbations within the long wavelength limit $\hbar\omega \ll \mu$.

As shown in the Supplemental Material (which includes references \cite{grimm1999optical,Ketterle99,Pitaevskii16,Giorgini98,Astrakharchik07}), in the regime of interest, both $\psi$ and $\theta$ are described by the Klein-Gordon Lagrangian. Equivalent to Eq.~(\ref{eq:Eff}) $\psi$ propagates at the speed of light, while $\theta$ propagates at the speed of sound in the BEC. Upon quantization, $\theta$ and $\delta \rho$ become noncommuting variables, though only the canonical momentum $\delta \rho = (\hbar / g_\mathrm{2d}) \partial_t \theta$ couples to the laser fluctuations (c.f.~Eq.~(\ref{eqn::Lint})). As the laser is operated at a frequency $\omega_L$ which is much higher than the frequencies at which the BEC is probed (i.e.~$\mu \ll \hbar \omega_L$), we can time average over the cycle period $2\pi/\omega_L$, which leads to the simplified interaction Lagrangian
\begin{equation}
\mathcal{L}_{\text{int}} ~=~ \alpha A_0^2\omega_L \delta\rho(t,\op{X}(t)) \partial_t\psi(t,0) ~.
\label{eqn::pertInt}
\end{equation}
Here we neglected the constant time-independent phase shift caused by the BEC bulk density, $\sim \rho_0 \partial_t \psi$, for simplicity and the zeroth order Stark potential can be canceled by using two laser beams with opposite detuning from the atomic resonance (see the following section and Supplement Material for details).

Since both fields, $\theta$ and $\delta \rho$, have the same spacetime dependence and the interaction Eq.~(\ref{eqn::pertInt}) is of the desired form Eq.~(\ref{eq:Lint}) our system is equivalent to the idealized field theory model, however, with the Wightman function Eq.~(\ref{Wightman}) evaluated for the canonical momentum, i.e. $\phi \equiv \delta \rho$. Before interaction with the BEC, the quantized field $\psi$ is simply electromagnetic noise (shot noise and phase noise). The interaction (\ref{eqn::pertInt}) generates correlations, as the laser phase samples quantum fluctuations in the BEC density along the interaction trajectory. To qualify as an observation of the analogue circular Unruh effect, one must be able to identify the characteristic trajectory dependence (\ref{TP}) from measurements made on the transmitted laser field.  In the following, we estimate the feasibility of making such an observation.

\begin{figure}[t]
\includegraphics[scale=.6]{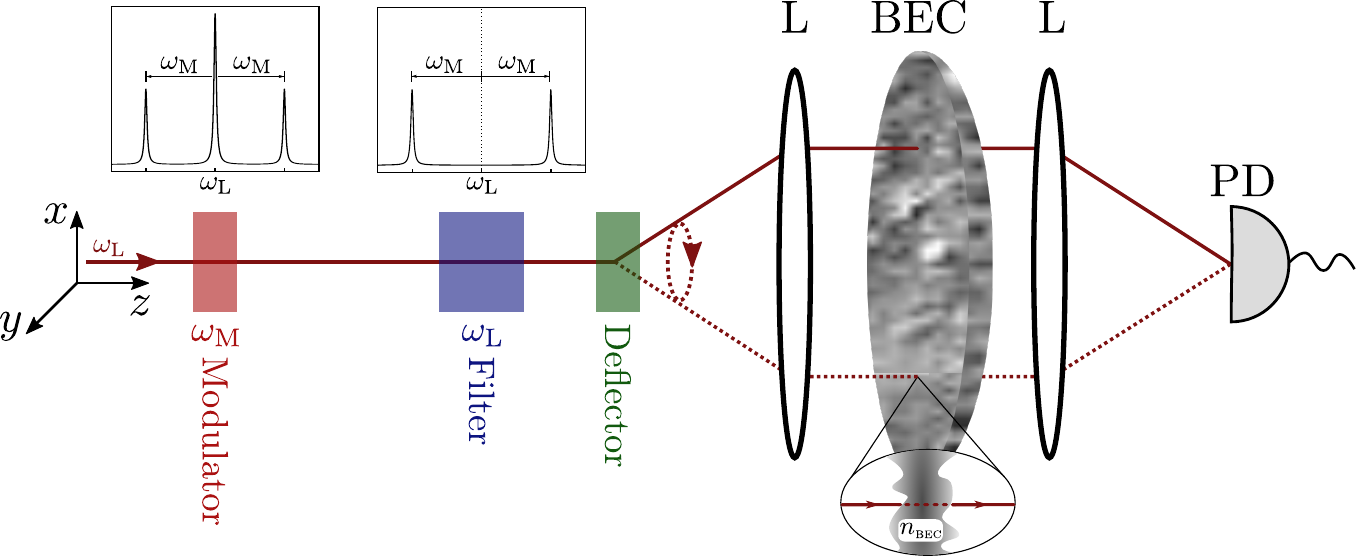}
\caption{Simplified schematic of our proposed experimental arrangement. An initial beam sharply peaked at frequency $\omega_L$ is modulated to create sidebands peaked at a pair of frequencies $\omega_L \pm \omega_M$ that are oppositely detuned from an atomic resonance. The central band is then filtered out. The two remaining sideband beams traverse the same path, intersecting a $2$d BEC along a circular trajectory in the BEC plane. Due to the opposite detuning, the sidebeams pick up opposite phase shifts when interacting with the BEC.}
\label{fig::interf}
\end{figure}

\textbf{\textit{Experimental Setup.}---}We will now describe and analyze a schematic interferometric setup (Figure~\ref{fig::interf}) that will allow to measure the fluctuations in a 2d quantum field, as discussed above.  
It is thereby crucial to reduce measurement-induced disturbances to the (quantum) minimum: a laser beam traversing a BEC creates a dipole potential, which will exert mechanical forces on the atoms. This can be counteracted by shining a second laser with opposite detuning along the identical path. The mechanical effects of the detector can then be compensated up to the (shot noise) intensity fluctuations of the two beams. Such an arrangement (Figure~\ref{fig::interf}) has in addition many other advantages: 
(i) the two laser beams experience opposite phase shifts (due to the opposite detuning from the atomic resonance)  which will allow better sensitivity; 
(ii) the two beams form an interferometer when a photodetector measures the beating between the two frequencies. The desired phase shift imprinted by the BEC quantum vacuum fluctuations can then be measured as a phase shift in the beating between the two opposite detuned laser frequencies;
(iii) the light beams go through the same path, hence any common disturbance will cancel when measuring their relative phase shift.

A deflector will then allow the beams (our detector) to interact with the BEC along a circular trajectory in the BEC plane. Continuous measurement of the phase quadrature then proceeds just as in heterodyne detection. The optimal detection regime can be obtained by constraining the photon absorption rate by BEC atoms, and optimizing the resulting signal-to-noise ratio (SNR) \cite{NDDetectors,AtomicClock}. 

In our system, the signal is characterized by the power spectrum of the phase difference between the two sidebands. It follows from the solution (\ref{psisol}) that the power spectrum for $\psi$ contains a contribution from the response function obtained by Fourier transforming the BEC field Wightman function (\ref{Wightman}) along the interaction trajectory. As long as the trajectory corresponds to a stationary worldline, the response function will be stationary, and can be compared with the response function for an Unruh-DeWitt detector.

We use the Unruh temperature given by (\ref{T_U}) to estimate the fluctuations sampled by the laser field as it interacts with the BEC. For later convenience, we define the dimensionless inverse analogue Unruh temperature of the circular trajectory as
\begin{equation}
\tilde{\beta} = \frac{\mu}{k_\mathrm{B} T_\mathrm{U}} = \frac{2 \pi c_s^2}{\gamma_s v^2} \frac{\mu}{\hbar \Omega_R} ~,
\label{T_U}
\end{equation}
where $\Omega_R = c_s / R$. Formula (\ref{T_U}) is obtained from (\ref{eq:T_unruh}) by using for $a$ the circular motion proper acceleration in the effective BEC Minkowski geometry and adjusting the energies to be defined with respect to the laboratory time, as is appropriate for the BEC\null. The actual effective temperature for circular motion includes an energy-dependent factor of order unity \cite{Good:2020hav,unruh98,bell83}, which we shall suppress here; a thorough analysis of this order unity factor in $(2+1)$ and $(3+1)$ dimensions is given in our companion paper \cite{UnruhDetectorLong}.

The resolution required to distinguish Unruh thermal from vacuum BEC fluctuations in the laser phase noise spectrum defines the scale of our signal; as shown in the Supplemental Material, the signal-to-noise ratio in the phononic regime $\hbar \omega \ll \mu$ for our proposed experiment is given by
\begin{align} \label{eq:dSNR_main}
\Delta SN \approx \sqrt{\frac{N \mathcal{B}}{2}} \chi \, \tilde{E}^2 \mathcal{F}(\tilde{\beta} \tilde{E}) \E^{- \left(r_0/\xi\right)^2 \, \tilde{E}^2 / 2} ~,
\end{align}
where $\tilde{E} = \hbar \omega / \mu$ is the BEC mode energy $\hbar \omega$ in units of the chemical potential, $\mathcal{F}(x) = (\E^x - 1)^{-1}$ is the Bose-Einstein distribution function, $N$ is the number of experimental realizations, $\mathcal{B}$ is the resolution bandwidth in units of the resolution bandwidth of the measurement $\mathcal{B}_\mathrm{m}$, and 
\begin{align} \label{eq:chi_SNR}
\chi &= \frac{3 \pi^2}{2} \left(\frac{\Gamma_\mathrm{sc} \E^{-\tilde{D}} \rho_0 \pi r_0^2}{2 \omega_L}\right) \left(\frac{\lambda_0}{\lambda_L}\right)^3 \left(\frac{m c^2}{\hbar \omega_L}\right) ~,
\end{align}
where $\Gamma_\mathrm{sc}$ is the photon scattering rate, $\tilde{D}$ is the off-resonance optical density, and $r_0$ is the beam width of the Gaussian laser beam. The exponential in Eq.~(\ref{eq:dSNR_main}) accounts for the averaging of fluctuations across the laser beam profile, leading to an additional suppression at high frequencies. 

\begin{figure}[t!]
\includegraphics[width=0.95\columnwidth]{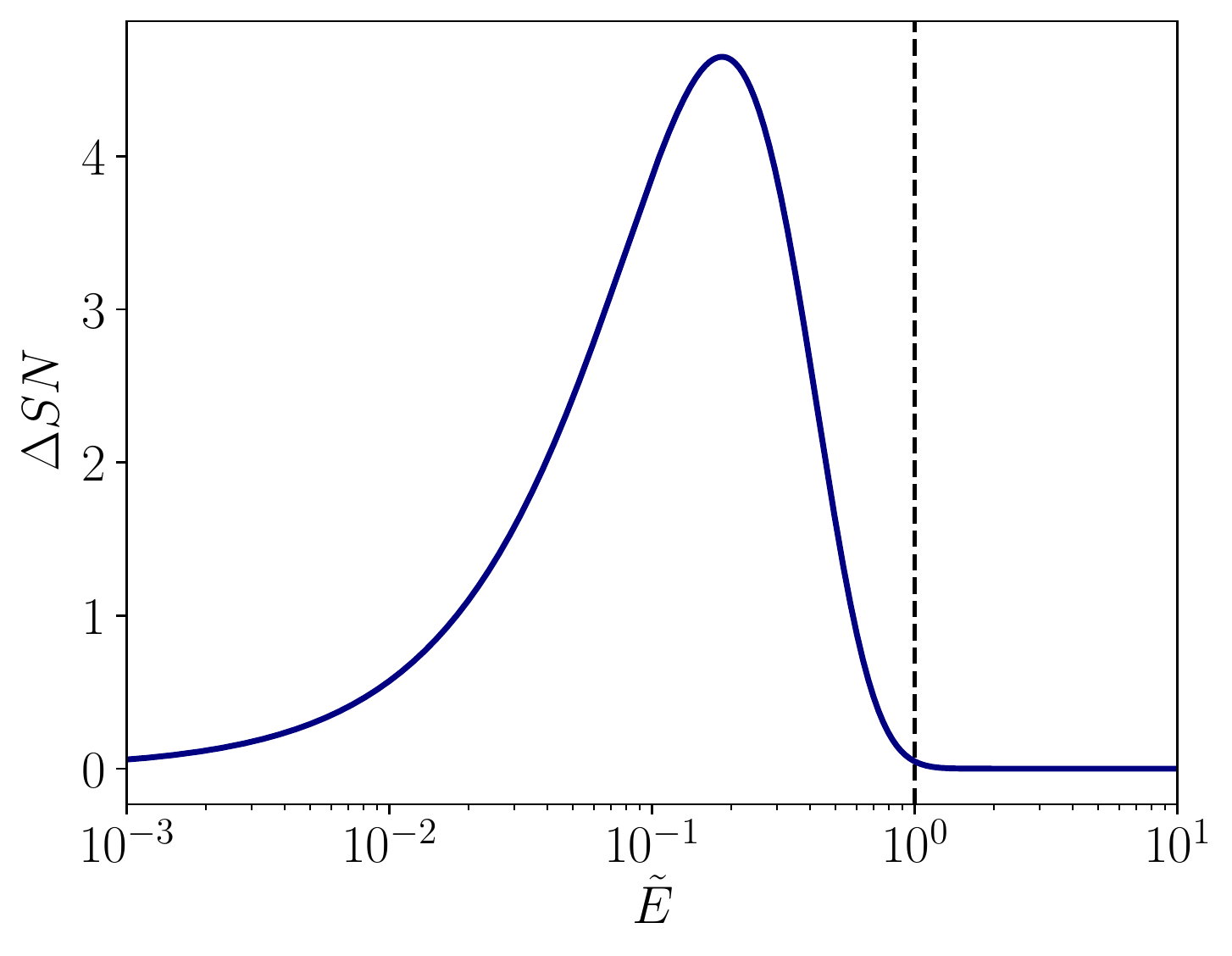}
\caption{Signal-to-noise ratio $\Delta SN$ as a function of dimensionless energy $\tilde{E} = \hbar \omega / \mu$ for $^{133}$Cs and $N=10^6$ experimental realizations. The experimental parameters chosen are the scattering rate $\Gamma_\mathrm{sc} \approx 0.1 \, \mathrm{Hz}$, beam width $r_0 = 3 \, \mu\mathrm{m}$, chemical potential $\mu \approx 2 \pi \hbar \, 9.5 \, \mathrm{Hz}$, healing length $\xi \approx 2 \, \mu\mathrm{m}$, density $\rho_0 = 10^3 \, \mu\mathrm{m}^{-2}$, scattering length $a_s = 25 \mathrm{pm}$, transverse confinement $a_\perp = 1 \mu\mathrm{m}$ resulting in a peak 3d density $\operatorname{max}[\rho_\mathrm{3d}] \approx 5.6 \cdot 10^{14} \, \mathrm{cm}^{-3}$ and speed of sound $c_s \approx 1.7\cdot 10^{2}\mu\mathrm{m}/\mathrm{s}$.  The observer trajectory radius $R = 10 \, \mu\mathrm{m} \approx 5 \xi$ and velocity $v = 0.95 c_s$ lead to $T_\mathrm{U} \approx 60 \, \mathrm{pK}$. Note that the signal vanishes for $\tilde{E} \to 0$, due to the suppression of density perturbations $\delta \rho$ at long wavelength. The signal to noise ratio within the whole phononic band (below the dashed black line) given by Eq.~(\ref{eq:dSNR_phonon_band}) for $\mathcal{B}_\mathrm{m} = 1 \, \mathrm{Hz}$ is $\overline{\Delta SN} \approx 5.8$.}
\label{fig:SNR_TU}
\end{figure}

Bounds on experimental parameters can readily be inferred from Eqs.~(\ref{T_U}), (\ref{eq:dSNR_main}), and (\ref{eq:chi_SNR}). First, continuous non-destructive measurement requires $\Gamma_\mathrm{sc} \ll 1$, limiting the power, detuning, and beam width of the laser beams. Second, $\mu / \hbar > \mathcal{B}_\mathrm{m}$ in order to have sufficient frequency resolution within the phononic (linearly dispersive) band. The measurement resolution bandwidth $\mathcal{B}_\mathrm{m}$ is limited by the lifetime of the BEC due to, e.g.,~technical heating or three-body collisions, and/or backaction of the laser. In particular, due to the finite size of the BEC, disturbances caused by the laser can reflect back from the edges, obscuring the desired behavior \cite{SoundProp}. Third, sufficiently high Unruh temperatures require $\hbar \Omega_R \approx \mu$, for which the observer radius $R$ has to be of the order of the healing length $\xi$ of the BEC. Combined, these requirements limit the laser parameters, the chemical potential $\mu$, and the dimensionless inverse Unruh temperature $\tilde{\beta}$. Note that heavy atomic species are favored since the SNR increases with the atomic mass $m$. Finally the signal and SNR can be maximized by increasing the density $\rho_0$ for which the scattering length $a_s$ must be decreased accordingly as to keep $\mu = \mathrm{const}$. The signal to noise ratio is shown in Fig.~(\ref{fig:SNR_TU}) for $\mathcal{B}=1$, and hence independent of the measurement bandwidth resolution. Note that for the results presented in Fig.~(\ref{fig:SNR_TU}) we, in particular, include the multilevel structure of atomic resonances in $^{133}$Cs (see Supplemental Material for details) which leads to a slight shift of the SNR predictions of order one from the theoretical prediction Eqs.~(\ref{eq:dSNR_main}) and (\ref{eq:dSNR_phonon_band}).

A lower bound on the observability can be given by considering the signal to noise ratio within the whole phononic band, given by
\begin{align} \label{eq:dSNR_phonon_band}
\overline{\Delta SN} &\approx \sqrt{\frac{N \mu}{4 \pi \hbar \mathcal{B}_\mathrm{m}}} \frac{\chi}{\tilde{\beta}^3} \mathcal{I}(\tilde{\beta}) ~.
\end{align}
The integral $\mathcal{I}(\tilde{\beta})$ is of order unity and explicitly given in the Supplementary Material. For $\mathcal{B}_\mathrm{m} = 1\, \mathrm{Hz}$ and the parameters presented in Fig.~(\ref{fig:SNR_TU}) we get $\overline{\Delta SN} \approx 5.8$.

\textbf{\textit{Conclusion.}---} We presented a new measurement scheme for the circular Unruh effect using a continuous probing field as a particle detector. The idealized model was then found to apply to the description of interacting fluctuations in a laser-coupled BEC system, for which preliminary estimates indicate that the proposed experimental implementation of the analogue cricular Unruh effect is within reach of current state-of-the-art cold-atom experiments. The proposed detection scheme is more generally useful for the field of quantum sensitive detectors for quantum fluids. The continuous measurement of fluctuations opens the door to new ways to probe quantum features not just of BECs, but of other laser-coupled systems such as superfluid helium. The backaction inherent to continuous measurements has not been addressed in this work; further investigation is required to determine how the effects of such backaction can be minimized. If necessary, the sensitivity could be enhanced by using a squeezed input state \cite{second}, as has been implemented by LIGO \cite{thirteen}.

\vspace*{0.5cm}
\noindent \textbf{Acknowledgements}: We would like to thank Pierre Verlot and Ralf Sch\"utzhold for stimulating discussions. This work originated at a June $2018$ Unruh effect workshop at the University of Nottingham, supported by FQXi (Mini-Grant FQXi-$MGB-1742$ ``Detecting Unruh Radiation'' to SW) and, in part, was made possible by UKRI STFC (UK) grant ST/S002227/1 "Quantum Sensors for Fundamental Physics". JL and WGU thank the organizers of the Relativistic Quantum
Information $2019$ School and Conference, Tainan, Taiwan, where
part of this work was done. JL and SW acknowledge partial support by Science and Technology
Facilities Council (Theory Consolidated Grant ST/P$000703/1$). WGU acknowledges support from the Hagler Institute for Advanced Studies at Texas A$\&$M University, the Canadian Institute for Advanced Research, the Natural Science and Engineering Research Council of Canada, the Helmholtz Association of German Research Centres, and the Alexander von Humbolt Foundation. SW
acknowledges support provided by the Leverhulme Research Leadership Award (RL-$2019-020$), the Royal Society University Research Fellowship (UF$120112$) and the Royal Society Enhancement Grant (RGF/EA/$180286$). SW and CG acknowledge the support provided by the Royal Society Enhancement Grant (RGF/EA/$181015$). SW and SE acknowledge support from the EPSRC Project Grant ($EP/P00637X/1$). JS and SE acknowledge support through the Wiener Wissenschafts- und TechnologieFonds (WWTF) project No $MA16-066$ (``SEQUEX''), and SE funding from the European Union's Horizon $2020$ research and innovation programme under the Marie Sklodowska-Curie grant agreement No $801110$ and the Austrian Federal Ministry of Education, Science and Research (BMBWF) from an ESQ fellowship. This letter reflects only the authors' view, the EU Agency is not responsible for any use that may be made of the information contained herein.

\bibliographystyle{apsrev4-1}
%

\clearpage
\begin{center}
\textbf{\large Supplemental Material}
\end{center}
\setcounter{equation}{0}
\setcounter{figure}{0}
\setcounter{table}{0}
\setcounter{page}{1}
\renewcommand{\theequation}{S\arabic{equation}}

\subsection{Polarizability volume}
The response of a dipole to a linearly polarized, external electromagnetic field is characterized by its polarizability $\alpha$. Approximating multi-level systems in the rotating wave approximation (RWA) with only one resonance frequency and neglecting the hyperfine-splitting, the frequency dependent polarizability volume can be expressed in terms of the resonance frequency $\omega_r$ and the linewidth $\Gamma$
\begin{equation}\label{eq:polarizabilitySimplified}
\alpha(\omega) = -\frac{6\pi c^3}{\omega_r^3}\frac{1}{\delta_0}+\I\frac{6\pi c^3}{\omega_r^3}\frac{1}{\delta_0^2}~,
\end{equation}
where $\delta_0=(\omega - \omega_r)/(\Gamma/2)$ is the detuning per half-linewidth, and we are assuming both that $|\delta_0|\gg 1$ and that the $|\omega - \omega_r|\ll \omega_r$ (see e.g.~\cite{grimm1999optical,Ketterle99,Pitaevskii16,Giorgini98,Astrakharchik07}). 

Equivalently, the polarizability (\ref{eq:polarizabilitySimplified}) can be used to define a fluctuating complex index of refraction ${n_\mathrm{BEC} = \sqrt{1 + \alpha \rho_\mathrm{3d}}}$, which can be expanded to linear order within the dilute gas approximation $\alpha \rho_\mathrm{3d} \ll 1$. The real part leads to a phase shift 
\begin{equation}
\varphi=\frac{2\pi(\operatorname{Re}\left[n_\text{BEC}\right]-1)\Delta z}{\lambda} ~,
\label{PS}
\end{equation}
whereas the imaginary part corresponds to absorption. Note that since $| \operatorname{Im}\left[\alpha\right]| = |\operatorname{Re}\left[\alpha\right]|/\delta_0$ absorption is highly suppressed for large detuning. Note that within the linear approximation ${n_\mathrm{BEC} \approx 1 + \alpha \rho_\mathrm{3d}}$ results are independent of the thickness $\Delta z$ and only depend on the 2d density $\rho_0 = \rho_\mathrm{3d} \Delta z$.

In the following analytic derivations, we will consistently use the assumptions as stated above and will neglect saturation effects, valid for low enough intensities, for simplicity. The results presented in the main text are calculated numerically, taking into account both the D1 and D2 resonance lines of $^{133}$Cs. The polarizability Eq.~(\ref{eq:polarizabilitySimplified}) for a single resonance is therefore, in case of linearly polarized light, replaced by 
\begin{align} \label{eq:polarizabilityD1D2}
\alpha = \alpha_\mathrm{D1}/3 + 2 \alpha_\mathrm{D2}/3 ~.
\end{align} 
For completeness, we take for $\alpha_\mathrm{D1/D2}$ predictions beyond the rotating wave approximation, valid for any value of the detuning (see e.g.~\cite{grimm1999optical}), but find good agreement with the simplified version Eq.~(\ref{eq:polarizabilitySimplified}).

\subsection{Linearisation of Single Laser Interaction}
In the following we derive the linearized BEC-light Lagrangian, connecting our setup to the idealized field theory given in the main text. We consider large detuning, such that we can neglect the complex part of the polarizability $\alpha$. The full Lagrangian is given by
\begin{align}
\mathcal{L} = \mathcal{L}_\mathrm{BEC} + \mathcal{L}_\mathrm{em} + \mathcal{L}_\mathrm{int} ~,
\end{align}
with the BEC, electromagnetic, and interaction Lagrangians provided in the main text.

The BEC Lagrangian is readily linearized following the standard procedure of expanding the complex field in the Madelung representation $\Phi = \sqrt{\rho_0 + \delta \rho} \, \operatorname{e}^{\I \theta}$ to second order in the small density perturbations $\delta \rho$ and phase gradients $\nabla \theta$. Considering for simplicity a constant background density $\rho_0 = N / V$, the equations of motion are given by
\begin{align} \label{eq:BEC_EOM}
\hbar\partial_t \delta \rho &= - \frac{\hbar^2}{m}\rho_0\nabla^2 \theta \notag \\
\hbar\partial_t \theta &=\frac{\hbar^2}{4 m \rho_0} \nabla ^2\delta \rho - g_{2\mathrm{d}}\delta \rho ~.
\end{align}
Upon canonical quantization, i.e.~imposing equal-time commutation relations $[\delta \hat{\rho}(\op{r}) \, \hat{\theta}(\op{r'})] = \I \delta(\op{r}-\op{r'})$, the equations of motion (\ref{eq:BEC_EOM}) can be mapped to the Bogoliubov equations for a condensate
\begin{align}
\I \hbar \partial_t \begin{pmatrix} B \\ B^{\dagger} \end{pmatrix} = \begin{pmatrix} - \frac{\hbar^2}{2 m} \nabla^2 + \mu & \mu \\ -\mu & \frac{\hbar^2}{2 m} \nabla^2 - \mu \end{pmatrix} \begin{pmatrix} B \\ B^{\dagger} \end{pmatrix} ~,
\end{align}
via the canonical transformation $B = \delta \hat{\rho}/2 \sqrt{\rho_0} + \I \sqrt{\rho_0} \hat{\theta}$. Therefore the standard diagonalization of the Bogoliubov Hamiltonian can be applied, leading to the modal expansion 
\begin{align}
&\hat{\theta}(t,\op{r}) = \frac{1}{2 \sqrt{V \rho_0}} \sum_k \sqrt{\frac{\epsilon_k}{E_k}} \left( \hat{b}_k \, \E^{-\I ( \nicefrac{\epsilon_k t}{\hbar} - \op{k} \op{r})} + \mathrm{H.c.} \right) \label{eq:BEC_theta} \\
&\delta \hat{\rho}(t,\op{r}) = \frac{\rho_0}{V} \sum_k \sqrt{\frac{E_k}{\epsilon_k}} \left( \hat{b}_k \, \E^{-\I (\nicefrac{\epsilon_k t}{\hbar} - \op{k} \op{r})} + \mathrm{H.c.} \right) \label{eq:BEC_rho}
\end{align}
for the density and phase fluctuations within the quasiparticle basis. Here $E_k = \hbar^2 k^2 / 2m$ and the dispersion relation is given by $\epsilon_k = \sqrt{E_k (E_k + 2 \mu)}$. In the long wavelength (phononic) regime, i.e.~neglecting the quantum pressure term $\sim \nabla^2 \delta \rho$ in Eq.~(\ref{eq:BEC_EOM}), we have $\epsilon_k \approx c_s \hbar k$ and each of the fields $\theta(t,\op{r})$,$\delta \rho(t,\op{r})$ obeys the massless Klein-Gordon equation.

For the electromagnetic field we calculate the derivative using the perturbed ansatz
\begin{align}
A(t,z) = A_0 \operatorname{cos}\left(\omega_L t - Kz + \psi(t,z)\right) ~.
\end{align}
Expanding the derivatives to second order in $\psi$ and time-averaging the fast oscillating terms as compared to the BEC dynamics, i.e.~$\sin(\omega_L t)^2\to\frac{1}{2}$, $\cos(\omega_L t)^2\to\frac{1}{2}$ and $\sin(\omega_L t)\cos(\omega_L t)\to0$, leads to
\begin{align}
\left(\partial_t A\right)^2 &= \frac{1}{2}A_0^2\left(\omega_L +\partial_t\psi\right)^2, \label{eqn::deltA} \\
\left(\partial_z A\right)^2 &= \frac{1}{2}A_0^2\left(-K +\partial_z\psi\right)^2. \label{eqn::delzA}
\end{align}
Since the terms linear in the first-order derivatives only contribute by a boundary term, the free electromagnetic Lagrangian can be effectively written as
\begin{equation} \label{eq:L_em_SM}
\lagrange{em}=\frac{A_0^2}{4}\left((\partial_t \psi)^2-(\partial_z \psi)^2\right) ~,
\end{equation}
which is, up to a constant rescaling of $\psi$, equivalent to the Klein-Gordon Lagrangian.

Similarly, the interaction Lagrangian
\begin{equation}
\mathcal{L}_{\text{int}} ~=~ \frac{\alpha}{2} \left(\partial_t A\right)^2\left|\Phi\right|^2,
\label{eqn::Lintsuppl}
\end{equation}
can be simplified by using (\ref{eqn::deltA}) to
\begin{equation}
\lagrange{int}=\frac{1}{4}\alpha A_0^2\left(\omega_L +\partial_t\psi\right)^2\left(\rho_0+\delta\rho\right).
\end{equation}
Hence, at leading order the coupling between the BEC and laser is
\begin{equation} \label{eq:L_int_SM}
\mathcal{L}_{\text{int}} = \frac{\alpha}{2} A_0^2\omega_L \delta \rho \partial_t\psi ~,
\end{equation}
The zeroth-order term in the laser phase fluctuations, responsible for the Stark-potential, is canceled by using two oppositely (red/blue) detuned laser beams with $(\alpha A_0^2 \omega_L^2)_\mathrm{blue} = -(\alpha A_0^2 \omega_L^2)_\mathrm{red}$. The constant phase shift caused by the bulk density also follows from (\ref{eqn::Lintsuppl}).

\subsection{Power spectral density and signal-to-noise ratio}
The measured observable is the power spectral density of the phase fluctuations in the laser, i.e.~the time-Fourier-Transform of the unequal-time phase-phase correlation function 
\begin{align} \label{eq:EM-phase-structureFaktor}
S_\psi(\omega) = \int_{-\infty}^{\infty} \! \mathrm{d}\tau ~ \langle \psi(\tau) \psi(0) \rangle ~ \E^{-\I \omega \tau} ~,
\end{align}
where $\tau = t - t'$ and we assumed time-translation invariance. From the solution to Eq.~(5) to linear order in the inverse fractional detunig (with the coupling $\epsilon$ defined by Eq.~(\ref{eq:L_int_SM})) after rescaling $\psi$ to bring the free electromagnetic Lagrangian Eq.~(\ref{eq:L_em_SM}) in canonical form we get
\begin{align} \label{eq:structure-factor-psi}
S_\psi(\omega) = \left( \frac{\pi \rho_0}{\lambda} \operatorname{Re}\left[\alpha\right] \right)^2 ~ S_{\delta \rho}(\omega) ~,
\end{align}
where $S_{\delta \rho}(\omega)$ is determined by the BEC density fluctuations $\delta \rho(t,\op{r})$ evaluated along the observer trajectory $\op{X}(t)$. Note that, for a fixed interaction point, Eq.~(\ref{eq:structure-factor-psi}) is equivalently obtained by linearizing the fluctuations of the refractive index $n_\mathrm{BEC}$. For non-inertial trajectories, however, $S_{\delta\rho}(\omega)$ will exhibit an excess in fluctuations caused by the Unruh effect.

In order to assess the feasibility to detect the signal caused by the Unruh effect we take a simplified approach, based on the expected approximately thermal response of an Unruh-DeWitt detector. We therefore estimate the feasibility for a stationary observer to detect an increase in fluctuations of the BEC caused by the Unruh temperature $T_U$. A more comprehensive theoretical description of the experiment is beyond the scope of this letter, and will be presented elsewhere.

For the experimental setup we consider a focused Gaussian laser beam with minimum beam width $r_0$, intensity $P_0$, frequency $\omega_L$ at the resonance frequency of the BEC. The initial beam is modulated with a microwave signal of frequency $\omega_M$ to populate the sidebands, with detuning $\delta_0 = \pm 2 \omega_M/\Gamma$ in units of the half-linewidth and power $P = M^2 P_0 / 4$ with modulation index $M$. The central band is subsequently filtered out.

The extracted shot-noise limited phase field
\begin{align}
\psi(t) = \psi_\mathrm{S}(t) + \delta\psi_\mathrm{SN}(t) ~.
\end{align} 
is averaged over the beam-size with the weight given by the Gaussian intensity profile
\begin{align} \label{eq:spatial_average_psi}
\overline{\psi}(t) = \int \mathrm{d}^2 r ~ \psi(r,t) ~ \frac{\E^{-\frac{r^2}{2 r_0^2}}}{2 \pi r_0^2} ~.
\end{align}
In the following we will suppress the overbar, denoting the spatial averaging, to shorten the notation. 

The signal $\psi_\mathrm{S}$, determined by Eq.~(\ref{eq:structure-factor-psi}), contains the information about the BEC fluctuations. Note that the red and blue detuned laser beams experience a phase shift $\pm \psi_\mathrm{S}$, respectively. Hence, assuming a sufficiently large photon number, the shot noise fluctuations can be approximated by a Gaussian with zero mean and variance
\begin{align} \label{eq:variance_shotNoise}
\langle \delta\psi_\mathrm{SN}(t) \delta\psi_\mathrm{SN}(t') \rangle = \frac{2 \hbar \omega_L}{M^2 P_0 \E^{-\tilde{D}}} \delta(t-t') = \sigma_\mathrm{SN}^2 \delta(t-t')~,
\end{align}
where $\tilde{D} = 2 \pi \rho_0 \operatorname{Im}{\alpha} / \lambda_L$ is the off-resonance optical density determining the transmitted power. All higher order cumulants vanish due to the Gaussian approximation.

The power spectral density is therefore given by
\begin{align}
\langle S_\psi(\omega) \rangle \approx S(\omega) + \sigma_\mathrm{SN}^2 ~,
\end{align}
where $S(\omega)$ for a thermal or vacuum state is determined through Eqs.~(\ref{eq:BEC_rho}),(\ref{eq:structure-factor-psi}),(\ref{eq:spatial_average_psi}), and for positive frequencies is given by
\begin{widetext}
\begin{align} \label{eq:signal_PSD}
S(\omega) = \left( \frac{\pi \rho_0 \operatorname{Re}[\alpha]}{\lambda_L} \right)^2 \, \frac{2 \pi m}{\hbar \rho_0} \frac{\mathcal{P}(\hbar \omega / \mu)^2}{\left(\hbar \omega / \mu\right)^2 - \mathcal{P}(\hbar \omega / \mu)} \mathcal{F}(\beta \hbar \omega) \E^{-\frac{2 m \mu r_0^2}{\hbar^2} \mathcal{P}(\hbar \omega / \mu)}~,
\end{align}
\end{widetext}
where $\mathcal{P}(x):=\sqrt{1+x^2}-1$, $\mathcal{F}(x) = (\E^{x}-1)^{-1}$ is the Bose-Einstein distribution function, and $\beta = (k_\mathrm{B} T)^{-1}$. The spatial averaging Eq.~(\ref{eq:spatial_average_psi}) leads to an additional suppression of the signal at high frequencies (last term in Eq.~(\ref{eq:signal_PSD})).

The variance of the mean $\bar{\sigma}_S$ of $\bar{S}$ for $N$ experimental realizations is
\begin{align}
\bar{\sigma}_S^2 = \frac{\sigma_S^2}{N \mathcal{B}} = \frac{\sigma_\mathrm{SN}^4}{N \mathcal{B}} \left[1 + 2 \frac{\bar{S}}{\sigma_\mathrm{SN}^2} + 2 \frac{\bar{S^2}}{\sigma_\mathrm{SN}^4} \right] ~.
\end{align}
Here, $\bar{S}$ is the signal averaged over the analysis resolution bandwidth $\mathcal{B}$, with $\mathcal{B}$ in units of the measurement resolution bandwidth $\mathcal{B}_\mathrm{m}$. We therefore define the signal-to-noise ratio (SNR)
\begin{align}
\mathrm{SN} := \frac{\bar{S}}{\bar{\sigma}_S} &= \sqrt{N \mathcal{B}} \left( \frac{\bar{S} / \sigma_\mathrm{SN}^2}{\sqrt{1 + 2 \frac{\bar{S}}{\sigma_\mathrm{SN}^2} + 2 \frac{\bar{S^2}}{\sigma_\mathrm{SN}^4}}} \right) \label{eq:SNR_full} \\
&\approx \sqrt{N \mathcal{B}} \frac{\bar{S}}{\sigma_\mathrm{SN}^2} ~, \label{eq:SNR_simple}
\end{align}
where we neglected the constant offset caused by the shot-noise and the second equality is valid for sufficiently small signals $S / \sigma_\mathrm{SN}^2 \ll 1$.

In order to assess the observability of the Unruh effect we consider the difference signal between a BEC vacuum state and a thermal state at the Unruh temperature $T_\mathrm{U}$. Since the signal $S(\omega)$ vanishes for zero temperature (see Eq.~(\ref{eq:signal_PSD})), we get
\begin{align} \label{eq:dSNR_phonon}
\Delta SN \approx \sqrt{\frac{N \mathcal{B}}{2}} \frac{\bar{S}_{T_\mathrm{U}}}{\sigma_\mathrm{SN}^2} ~,
\end{align}
where we restrict our attention to the approximate form Eq.~(\ref{eq:SNR_simple}) for simplicity.

\begin{table}[b!]
{\setlength{\tabcolsep}{1.5em}
\renewcommand{\arraystretch}{1.5}
\begin{tabular}{ | c | c |}
\hline
inverse healing length & $\xi^{-1} \sim \sqrt{\frac{a_s}{a_\perp} \rho_{2d}}$ \\
\hline
chemical potential & $\mu \sim \xi^{-2} m^{-1}$ \\
\hline
inverse Unruh temperature & $\beta_{T_U} \sim \xi R m$ \\
dimensionless & $\tilde{\beta} \sim R \xi^{-1}$ \\
\hline
$\Delta SN$ prefactor & $\chi \sim \frac{a_\perp}{a_s} \xi^{-2} m$ \\
 \hline
\end{tabular}}
\caption{List of important scaling relations.}
\label{tb:scalings}
\end{table}

For the remainder we will consider the relevant phononic limit $\hbar \omega < \mu$ for simplicity. The explicit results reported in the main text are calculated numerically for the full model, using Eqs.~(\ref{eq:polarizabilityD1D2}), (\ref{eq:signal_PSD}), and (\ref{eq:SNR_full}). We find adequate accordance with the simplified analytic predictions, the full model showing small deviations of order one.

In the phononic regime $\hbar \omega < \mu$ and for large detuning $\delta_0 \gg 1$, Eq.~(\ref{eq:dSNR_phonon}) simplifies to
\begin{align}
\Delta SN \approx \sqrt{\frac{N \mathcal{B}}{2}} \chi \, \tilde{E}^2 \mathcal{F}(\tilde{\beta} \tilde{E}) \E^{- \left(r_0/\xi\right)^2 \, \tilde{E}^2 / 2} ~.
\end{align}
Here, $\tilde{E} = \hbar \omega / \mu$ is the BEC mode energy $\hbar \omega$ in units of the chemical potential and $\tilde{\beta} = \beta \mu$ is the dimensionless inverse temperature. The dimensionless constant 
\begin{align}
\chi &= \frac{3 \pi^2}{2} \left(\frac{\Gamma_\mathrm{sc} \E^{-\tilde{D}} \rho_0 \pi r_0^2}{2 \omega_L}\right) \left(\frac{\lambda_0}{\lambda_L}\right)^3 \left(\frac{m c^2}{\hbar \omega_L}\right)
\end{align}
is determined through the atom-light interaction parameters and the photon scattering rate ${\Gamma_\mathrm{sc}=\frac{1}{\hbar c \pi r_0^2} \operatorname{Im}\left[\alpha\right] M^2 P_0}$, evaluated at the peak intensity of the Gaussian laser beam. Note that we used Eq.~(\ref{eq:polarizabilitySimplified}) to relate the real and imaginary part of the polarizability. Continuous non-destructive measurement of the BEC requires $\Gamma_\mathrm{sc} \ll 1$.

Neglecting the high frequency cutoff, which is small within the phononic regime, the frequency dependence of the SNR is determined by the function $f_S(\tilde{E},\tilde{\beta}) = (\tilde{\beta} \tilde{E})^2 \mathcal{F}(\tilde{\beta} \tilde{E}) / \tilde{\beta}^2$. We therefore get the dependence of the SNR on the system parameters $\Delta SN \equiv \Delta SN[\rho_0,m,\tilde{\beta}]$. Keeping the remaining parameters fixed, we find: (i) increasing (decreasing) the density $\rho_0$ shifts the SNR to lower (higher) frequencies, (ii) increasing (decreasing) the atomic mass $m$ shifts the SNR to higher (lower) values, and (iii) increasing (decreasing) $\tilde{\beta}$, i.e.~increasing (decreasing) Unruh temperature, self-similarly shifts the SNR along a parabola to lower (higher) frequencies. The maximum SNR is reached at $\tilde{E}_\mathrm{max} \approx 1.7/\tilde{\beta}$ with $f_S(\tilde{E}_\mathrm{max},\tilde{\beta}) \approx 0.65/\tilde{\beta}^2$. Note that for $\tilde{\beta} \leq 1.7$ the maximum is approximately at $\tilde{E}_\mathrm{max} \approx 1$ with $f_S(\tilde{E}_\mathrm{max},\tilde{\beta}) \approx 0.6 \mathcal{F}(\tilde{\beta})$, where the numerical factor stems from the suppression at high frequencies in the full expression Eq.~(\ref{eq:signal_PSD}) valid beyond the phononic regime. Important scaling relations with with relevant experimental parameters are summarized in table (\ref{tb:scalings}).

A lower bound on the observability of the signal can be given by considering the SNR within the whole phononic band, i.e.~$\mathcal{B} = \mu / (2 \pi \hbar \mathcal{B}_\mathrm{m})$. This estimates the ability to detect the increase in power due to access of fluctuations caused by the BEC within the regime where the analogue Unruh effect is expected to be valid. Neglecting again the high frequency cutoff, valid for sufficiently small beam width $r_0 < \xi$, we get the closed equation
\begin{align}
\overline{\Delta SN} &\approx \sqrt{\frac{N \mu}{4 \pi \hbar \mathcal{B}_\mathrm{m}}} \frac{\chi}{\tilde{\beta}^3} \mathcal{I}(\tilde{\beta}) \\ 
&\approx \chi \sqrt{\frac{N \mu}{4 \pi \hbar \mathcal{B}_\mathrm{m}}} \left( \frac{\gamma_s v^2 \hbar \Omega_R}{2 \pi c_s^2 \mu} \right)^3 \mathcal{I}(\tilde{\beta})~.
\end{align}
For the second equality we inserted the Unruh temperature from the main text, with $\Omega_R = c_s/R$. From table (\ref{tb:scalings}) we find $\overline{\Delta SN} \sim \sqrt{m} (a_\perp/a_s) R^{-3}$, showing the dependence on the independent BEC and observer trajectory parameters. The integral is given by
${\mathcal{I}(\tilde{\beta}) = -(\tilde{\beta}^3 / 3 ) \! - \! \tilde{\beta}^2 \operatorname{Li}_1(\E^{\tilde{\beta}}) \! + \! 2 \tilde{\beta} \operatorname{Li}_2(\E^{\tilde{\beta}}) \! - \! 2 \operatorname{Li}_3(\E^{\tilde{\beta}}) \! + \! 2 \zeta(3)}$,
with the polylogarithm functions $\operatorname{Li}_s$ and the zeta-function $\zeta$, and rapidly approaches its limit $\lim\limits_{\tilde{\beta} \to \infty} I(\tilde{\beta}) = 2 \zeta(3)$ for $\tilde{\beta} \gtrsim 5$.

\end{document}